**Dynamical System Modeling Of Immune Reconstitution Following Allogeneic Stem Cell Transplantation Identifies Patients At Risk For Adverse Outcomes.**

Amir A. Toor [1], Roy T. Sabo [2], Catherine H. Roberts [1], Bonny L. Moore [1], Salman R. Salman [1], Allison F. Scalora [1], May T. Aziz [3], Ali S Shubar Ali [1], Charles E. Hall [1], Jeremy Meier [1], Radhika M Thorn [1], Elaine Wang [1], Shiyu Song [4], Kristin Miller [5], Kathryn Rizzo [6], William B. Clark [1], John M. McCarty [1], Harold M. Chung [1], Masoud H. Manjili [7] and Michael C. Neale [8].

[1] Bone Marrow Transplant Program, Massey Cancer Center, Department of Internal Medicine, [2] Department of Biostatistics, [3] Department of Pharmacy and Therapeutics, [4] Department of Radiation Oncology, [5] Department of Internal Medicine, [6] Department of Pathology, [7] Department of Microbiology & Immunology and the [8] Departments of Psychiatry & Human Genetics Virginia Commonwealth University, Richmond, Virginia.

Address correspondence to Amir A. Toor MD

Associate Professor of Medicine,

Massey Cancer Center, Virginia Commonwealth University,

1300 E Marshall St, PO box 980157

Richmond, VA 23298-0157

Ph: 804-828-4360.

Fax: 804-828-7825.

E-mail: *atoor@mcvh-vcu.edu*

**Abstract**

Systems that evolve over time and follow mathematical laws as they do so, are called *dynamical systems*. Lymphocyte recovery and clinical outcomes in 41 allograft recipients conditioned using anti-thymocyte globulin (ATG) and 4.5 Gray total-body-irradiation were studied to determine if immune reconstitution could be described as a dynamical system. Survival, relapse, and graft vs. host disease (GVHD) were not significantly different in two cohorts of patients receiving different doses of ATG. However, donor-derived CD3+ (*dd*CD3) cell reconstitution was superior in the lower ATG dose cohort, and there were fewer instances of donor lymphocyte infusion (DLI). Lymphoid recovery was plotted in each individual over time and demonstrated one of three sigmoid growth patterns; Pattern A (n=15), had rapid growth with high lymphocyte counts, pattern B (n=14), slower growth with intermediate recovery and pattern C, poor lymphocyte reconstitution (n=10). There was a significant association between lymphocyte recovery patterns and both the rate of change of *dd*CD3 at day 30 post-SCT and the clinical outcomes. GVHD was observed more frequently with pattern A; relapse and DLI more so with pattern C, with a consequent survival advantage in patients with patterns A and B. We conclude that evaluating immune reconstitution following SCT as a dynamical system may differentiate patients at risk of adverse outcomes and allow early intervention to modulate that risk.

**Introduction.**

Allogeneic stem cell transplantation (SCT) results in widely disparate outcomes in individual transplant recipients, regardless of uniformity of histocompatibility criteria applied in donor selection [1, 2, 3] and therapeutic interventions [4, 5, 6] utilized for pre transplant conditioning regimens and graft versus host disease (GVHD) prophylaxis. Standard methodology is to examine clinical outcomes using statistical analytic techniques based on probability theory [7, 8]. These analytic techniques are useful in determining odds of clinical outcomes in populations of patients transplanted using uniform conditioning regimens, but are inadequate for determining the course an individual might take following SCT. This is because of the underlying assumption that, within the constraints of certain critical parameters, such as donor type or HLA matching, the probability distribution of these clinical outcomes is essentially random. For example, in patients allografted for myelodysplastic syndrome, the addition of antithymocyte globulin (ATG) to the conditioning regimen reduced the *likelihood* of developing severe acute GVHD [9]. Similarly, higher levels of donor T cell chimerism at day-30 following reduced-intensity (busulfan+fludarabine) conditioning resulted in a lower *probability* of relapse following SCT [10]. Such population-based estimates may then be used to guide clinical decision-making at various time points during the transplant process, but individual patients undergoing SCT remain at risk for competing causes of adverse outcomes.

To improve outcome predictions in individuals undergoing SCT in real-time, a closer examination of the biology of transplantation is necessary, specifically, the immunobiology of transplantation. In patients transplanted for management of hematological malignancy treatment failure is often attributable to either the development of life-threatening GVHD or the lack of graft versus malignancy (GVM) effects [11, 12]. These are, in turn influenced by the immune reconstitution following transplantation [13, 14, 15, 16]. Hypothetically, repertoire complexity emerging in the reconstituting cell populations may be driven, in part, by the disparity in minor histocompatibility antigens encountered in each instance of SCT [17, 18]. It has been demonstrated that HLA-matched donor-recipient pairs have extensive variation in their exomes, [19] and *in silico* this translates to a massive array of minor histocompatibility antigens, which may be presented by the HLA in the recipients [20]. This minor histocompatibility antigen difference may be considered an alloreactivity potential between donors and recipients, and appears to be very large when the number of variant peptide-HLA complexes is accounted for in each pair. Further, the binding affinity distribution of these peptides to the HLA appears to mirror the complex T cell clonal frequency observed in the T cell repertoire in SCT donors and in recipients [21]. These findings, suggest that clinical outcomes related to T cell alloreactivity following SCT may not primarily be stochastic in nature, rather when taking a quantitative perspective may be better considered mathematically to account for the large alloreactivity potential between SCT donors and recipients. Systems that follow mathematically defined laws are common in the natural world and are called *dynamical systems* [22, 23, 24, 25]. A dynamical system is defined as any *iterating* physical system, which *evolves* over time in a manner such that *future states* of the system are predicated on all the *preceding states*, and the evolution of the system can be modeled using ordinary differential equations. Such systems will behave either in *linear* (such as a mass accelerating in proportion to the force applied) or *non-linear* fashion (such as population growth constrained by environmental pressures). In the context of SCT this implies that if considered as a system of interacting donor T cell clones and recipient antigens, immune reconstitution following SCT is a dynamic, evolving process, which can be modeled mathematically and allow more precise determination of the odds of clinical outcomes such as engraftment, GVHD and survival in an individual [26]. A feature of dynamical systems is that early events in the system set the trajectory of the series of events to follow, and thus determine the eventual outcome. In certain non-linear dynamical systems, this implies that small variations in the early state of the system can produce large measurable effects late in the evolution of this system. In SCT a large body of literature now supports the notion that both early interventions [27, 28, 29], as well as the magnitude of immune reconstitution early in the course of SCT impacts late clinical outcomes [13, 30, 31], supporting the validity of considering SCT as a nonlinear dynamical system.

In this paper the results of a prospective randomized phase-2 clinical trial are reported, and considered in the light of immune reconstitution kinetics modeled as a dynamical system. The trial investigated two doses of rabbit anti-thymocyte globulin (ATG; Thymoglobulin, Sanofi-Aventis), in patients undergoing reduced intensity conditioning (RIC). The trial was designed to determine the optimal dosing of ATG to be given in combination with reduced intensity total body irradiation (TBI) to ensure adequate engraftment. The optimal ATG dose should allow adequate immune reconstitution conferring protection from relapse with a lower risk of GVHD, and that regimen would eventually serve as a comparator for conventional RIC regimens in phase III trials. The clinical outcomes from this trial are analyzed with an underlying assumption, that each transplant represents an example of a dynamical system. Each donor-recipient-pair is comprised of a unique set of recipient alloantigens and a set of donor derived immune effectors, interacting over time following the transplant event, in this instance modulated by two different ATG doses. The focus of the work presented here has been on total lymphoid and T cell reconstitution and the resulting clinical effect. We demonstrate that lymphoid reconstitution following transplantation follows the ubiquitous quantitative rules describing constrained growth, i.e., it occurs as a logistic function of time.

**Methods.**

*Patients*

Between 2009 and 2013, 41 patients were enrolled on a prospective randomized phase II clinical trial, approved by the institutional review board at Virginia Commonwealth University (ClinicalTrials.gov Identifier: NCT00709592). To be eligible, patients had to be ≤70 years of age, with recurrent or high-risk hematological malignancy. Patients younger than 50 years had to be ineligible for conventional myeloablative conditioning because of comorbidity. The patients were required to have a 7/8 or 8/8 locus matched related (MRD) or unrelated donor (URD), with high-resolution typing performed for HLA-A, B, C and DRB1. Two different doses of ATG were tested in the two arms of this trial, and patients were stratified for donor type (MRD or URD) and disease status (CR1 or >CR1) at the time of randomization. Patient characteristics are given in Table 1. Patient follow up is updated as of October 2014.

*Rabbit ATG & Low-dose Total Body Irradiation Conditioning Regimen*

Patients were randomized between two different doses of rabbit-anti-thymocyte globulin, either ATG 2.5 or 1.7 mg/kg, adjusted ideal body weight per day given intravenously from day −9 to −7 (ATG 7.5 or ATG 5.1 cohorts), followed by total body irradiation (TBI) to a total dose of 4.5 Gray, delivered in three 1.5 Gray fractions, administered twice on day −1, with the final dose on day 0 (Supplementary Figure 1). Uniform supportive care, ATG premedication and antimicrobial prophylaxis were administered (Supplementary material). GVHD prophylaxis was with tacrolimus given orally from day −2 with taper commencing around 12 weeks post transplant, in an initial cohort of patients (n=24), and according to donor derived T cell counts in the remaining patients. Mycophenolate mofetil was given orally at a dose of 15 mg/kg twice daily from day 0 to 30. G-CSF was given at a dose of 5 µg/kg/d from day 4 until myeloid engraftment. Blood stem cells were collected using G-CSF 10 µg/kg/d given subcutaneously on day 1 through 5. Escalating dose DLI was permitted beyond 8 weeks post SCT for the management of declining or persistent mixed chimerism and for disease progression.

*Lymphoid reconstitution following SCT*

Absolute lymphocyte counts ($\mu L^{-1}$) (ALC) measured at multiple time points as a part of routine clinical care of the patients following SCT were plotted over time and analyzed using GraphPad Prism version 6.00 for Windows (GraphPad Software, La Jolla California USA). These data were examined for trends over time and further analyzed as a logistic function of time utilizing the equation of the general form, $N_t = N_0 + (K-N_0)/(1+10^{(a-t)R})$, (adopted from Prism v 6.00). In this equation, $N_0$ represents the lymphocyte count at the beginning $K$ is the lymphocyte count at steady state; $Nt$, is the lymphocyte count at time $t$ following transplant; $a$, is the time at which growth rate is maximal and an inflection point is observed in the logistic curve, and $R$ is the growth rate of the population. In order to elucidate the initial growth followed by exponential expansion of the lymphocytes, and to eliminate the effect of extraneous therapy on lymphocyte counts (e.g., corticosteroids for GVHD, chemotherapy for relapse, CMV reactivation and therapy) these parameters were determined only for the time post transplant before these clinical events had transpired. ALC following DLI were not considered as a part of this analysis.

*T cell engraftment analysis*

Donor engraftment was measured using chimerism analyses performed at approximately 4, 8, 12, and 24 weeks following transplant on immunomagnetically separated granulocytes, and total T cells as previously described [13]. *Donor-derived* CD3+ T cell count (*dd*CD3) was calculated by multiplying the T cell chimerism (% recipient DNA, expressed as a fraction) with the absolute blood CD3+ T cell count obtained simultaneously. The resulting absolute recipient-derived CD3+ T cell count was then subtracted from the total absolute CD3+ cell count, to obtain a *donor-derived* CD3+ T cell count. When available, the calculated *dd*CD3 values for the first three to six months were plotted over time for each patient as a polynomial (cubic) function of time, ($y = ax^3+bx^2+cx+d$ using Microsoft Excel software. This was done with the assumption that the *dd*CD3 on day 0 of SCT will be zero. From the resulting polynomial equations for each patient, the differential equation describing the rate of change of *dd*CD3 at any time point along the curve for each patient was determined ($dy/dx = 3ax^2+2bx+c$). The derivative, i.e. rate of change of *dd*CD3 at day 30 and day 45 was then calculated for each individual using Wolfram/Alpha Online Derivative Solver. These early points were chosen as they represented the earliest clinically important measured *dd*CD3 value available and because these allowed determination of the T cell expansion rate in the time period immediately following the cessation of MMF. The derivative calculations were also confirmed by manual calculations.

*Study Design and Statistical analysis*

For this analysis overall survival was taken from the day following transplant to the day of death. GVHD was classified according to consensus criteria. Acute GVHD was graded according to the Glucksberg criteria. GVHD observations were censored if this complication developed after DLI. Disease specific criteria were used for diagnosing relapse or progression. Survival and event-free survival curves (relapse-fatality) for the 5.1- and 7.5-level dose groups are plotted with Kaplan-Meier curves and are compared using the log-rank test. The incidence curves for relapse, acute GVHD, chronic GVHD, DLI use (accounting for the competing risk of mortality) are plotted and compared between dose levels using Grey's test. Wilcoxon rank sum tests are used to compare median percentage chimerism and *dd*CD3 cell count between dose groups at different times following SCT; this non-parametric test is used because these measures exhibit highly non-symmetric distributions. The R statistical software (version 2.15) was used for all time-to-event analyses, with the *survival* package used for survival curves, and the *cmprsk* package used for all competing risk curves.

The rate of change in *dd*CD3 cell count at day 30 ($dx/dy(30)$) and day 45 ($dx/dy(45)$) following SCT is summarized with medians and inter-quartile ranges for each level (Yes/No) of the following clinical outcomes: GVHD, Relapse, Survival and DLI use. The Wilcoxon rank sum test is used to compare medians in $dx/dy(30)$ and $dx/dy(45)$ between the levels of those health outcomes and other variables studied. A 5% significance level is used for purposes of testing null hypotheses that these values are equal across the two levels of the clinical outcomes vs. two-sided alternative hypotheses. The Kruskal-Wallis test is used to compare monocyte count medians between lymphocyte recovery patterns. A 5% significance level is used for purposes of testing null hypotheses, and a Bonferroni-adjusted level of 0.1067 is used in cases of multiple comparisons between logistic characteristic patterns. The NPAR1WAY procedure is used in the SAS Statistical Software (version 9.4, Cary, NC, USA) for all analyses. The Pearson correlation coefficient was used to measure the relationship between logistic growth patterns in lymphocyte recovery and CD3+ dose, CD34+ dose, and tacrolimus level area under the curve.

**Results.**

*Clinical outcomes with ATG+TBI conditioning*

This study was designed to examine the relative impact of two ATG doses in a RIC regimen on immune reconstitution and subsequent clinical outcomes, with the aim of improving survival by limiting fatal GVHD while retaining GVM effects. To date, there is no significant difference in the survival between the two study arms (Log-Rank P = 0.53) (Figure 1A). The 1- and 2-year survival rates in the ATG 5.1 cohort are both 71.3% (SE = 1.05%), while those in the 7.5 cohort are 90.9% (SE = 0.39%) and 62.4% (SE = 1.19%), respectively. The conditioning regimen was tolerated well with no day 100 transplant related mortality observed. The treatment related mortality rate and relapse rate were not significantly different between the two dose groups (Grey's test P=0.82 & 0.62 respectively) (Figure 1B). The 1- and 2-year relapse rates in the ATG 5.1 cohort are 21.8% (SE = 0.79%) and 28.0% (SE = 1.06%), respectively, and in the ATG 7.5 cohort are 40.9% (SE = 1.19%) and 50.0% (SE = 1.22%), respectively. Event free survival was likewise, comparable between the two arms (P=0.76) (Figure 1C).

Accounting for the competing risk of mortality, acute GVHD rate in the ATG 5.1 cohort is 27.2% (SE = 0.96%), and in the 7.5 dose group it was 4.5% (SE = 0.21%) (P=0.039). Seven patients developed grade II-IV acute GVHD (6 grade III-IV); six were in the ATG 5.1 arm; one of those patients did not receive tacrolimus after day 10 due to neurotoxicity and was switched to sirolimus. All but one patient developed acute GVHD after immunosuppression was withdrawn, with the median time to maximum grade of acute GVHD being day 130 post-SCT. Similarly, accounting for the competing hazard of mortality, the 1- and 2-year chronic GVHD rates in the ATG 5.1 cohort are both 23.8% (SE = 0.94%), and in the ATG 7.5 group are 31.8% (SE = 1.05%) and 36.8% (SE = 1.16%), respectively (P=0.37) (Figure 1D). Chronic GVHD was diagnosed at a median 172 days post-SCT in 12 patients (4 in the ATG 5.1 arm) and was moderate to severe in 8 patients.

The need for post-transplant cellular therapy intervention, DLI (administered for either mixed chimerism or persistent disease or relapse) was greater in the ATG 7.5 arm. Accounting for the competing risk of mortality, the 1- and 2-year DLI rates in the ATG 5.1 arm are both 8.9% (SE = 0.38%), while in the ATG 7.5 arm it was 40.9% (SE = 1.12%) and 45.5% (SE = 1.21%), respectively (Grey's test P=0.072) (Figure 1E). Median time to DLI was 237 days and a median 3 infusions were administered, with a CD3 cell dose ranging from $10^6$ to $5\times10^7$ CD3+ cells/kg. Nine of the thirteen patients receiving DLI developed GVHD following DLI, this was fatal in two cases.

*Engraftment*

To determine the dose effect of ATG on T cell and myeloid engraftment in this RIC regimen, T cell chimerism and absolute T cell count recovery, as well as granulocyte chimerism were compared between the two arms. T cell chimerism was mixed donor-recipient in a number of patients, with a larger recipient derived component in the ATG 7.5 cohort compared with the ATG 5.1 cohort except at 9 months (where both medians are 0). These differences are significant at 12 weeks (p-value = 0.02), and approach significance at 6, 9 and 12 months (Table 2). Correspondingly, the *dd*CD3 cell counts, are higher in the ATG 5.1 cohort than in the ATG 7.5 cohort at every time point (Supplementary figure 2), being significant at 12 weeks (p-value = 0.03), and approaching significance at 4 and 8 weeks and at 9 months post-SCT (Table 2). Myeloid engraftment was completely donor-derived in all but 3 patients in the ATG 7.5 cohort. Two other patients with MDS had persistent disease with mixed myeloid chimerism, which resolved with withdrawal of immunosuppression in one and with administration of DLI in the other. These results demonstrate that hematopoietic engraftment is superior in the ATG 5.1 cohort.

*Lymphoid reconstitution kinetics*

In order to study the influence of the rate of immune reconstitution on clinical outcomes in individual patients, absolute lymphocyte counts (ALC, $\mu L^{-1}$) for each individual were plotted over time to model immune reconstitution kinetics. Similar to population growth models, lymphocyte recovery following SCT occurred as a logistic function of time. In a number of patients, 2 discernable periods of exponential increase in ALC were observed, one following engraftment, and another period following cessation of MMF (Figure 2A, and Supplementary Figure 3). These growth periods were each followed by a plateau, representing the steady state immune cell recovery at that phase of SCT, with relatively stable lymphocyte counts in the absence of clinical events until withdrawal of tacrolimus when greater variability was observed, and a trend of a further increase in ALC was frequently observed. On examination of individual lymphocyte recovery plots, 3 general patterns of lymphoid recovery were observed over time (Figure 2B); sigmoid growth with early, rapid lymphoid expansion and a high steady state level (pattern A; ALC >1000 $\mu L^{-1}$, inflection point <60 days post SCT; n=15); a lower steady state level arrived at slowly (pattern B; ALC ≥500-1000 μL$^{-1}$, inflection point >60 days; n=14) and poor ALC

recovery with minimal to no discernable sigmoid growth kinetics (pattern C; ALC <500 μL$^{-1}$; n=10). Two

patients from the ATG 7.5 cohort with good lymphoid recovery (pattern A) but predominantly recipient

T cell chimerism were excluded from this analysis. Patients developing complications of therapy such as relapse, viral reactivation (CMV or EBV) and GVHD requiring corticosteroid therapy, had significant departures from the sigmoidal curve generally after reaching a plateau (Supplementary Figure 3).

Logistic curve fitting demonstrated patients with pattern A demonstrated the double exponential

expansion periods, (median R$^2$, 0.90 and 0.89 respectively, n=15), with median growth exponent ($R$ in

the logistic equation) values of 0.94 and 0.18 for the two periods respectively; pattern B patients

similarly demonstrated logistic growth (median R$^2$ 0.83, n=14), albeit with a much slower growth rate of

0.04.

Logistic patterning was significantly associated with all clinical outcomes, including survival in 67% vs.

86% vs. 30% patients in the A, B and C groups respectively (Exact test P value= 0.0189); relapse in 33, 29

and 90% (P=0.0053); cumulative GVHD in 67, 43 and 10% (P=0.0198) and DLI use in 13, 21 and 70%

patients respectively (P=0.0069). The patients with pattern C had a particularly poor outcome in terms of survival, relapse and need for DLI.

*Determinants of Logistic Patterning*

To identify factors influencing logistic growth patterns observed in lymphocyte recovery; ATG dose,

donor type and age, CD3+ and CD34+ cell doses infused, tacrolimus level area under the curve for days

15, 30 and 60 were reviewed and did not demonstrate a significant association with the ALC curve logistic patterns recorded. Reasoning that the lymphocyte growth exponent, $R$, for T cells will be influenced by antigen presentation prior to the period of exponential growth, the monocyte recovery curves following SCT were plotted for each individual and overlaid with the T cell expansion curve.

Monocyte count was used as a surrogate for antigen presenting cells, since dendritic cells would partly arise from this population.  Following engraftment, there was correspondence in the lymphocyte and monocyte growth curves, both demonstrating logistic dynamics (Figure 3A, Supplementary Figure 4).

Further there was a strong association between the maximal circulating monocyte counts (μl$^{-1}$)

recorded between day 12 and 16 (peri-engraftment) and the logistic patterns observed in lymphocyte recovery over the first 3 months post-transplant; these values were 600, 350 and 200 respectively in patients with lymphocyte recovery patterns A, B and C (p value=0.0002). While this correspondence may be a reflection of robust hematopoietic engraftment, it is reasonable to hypothesize that the robust monocyte recovery post engraftment is associated with greater likelihood of recipient (and microbial) antigen presentation in the earliest days following SCT. This would result in higher values of $R$, and faster lymphocyte expansion, potentially explaining the greater incidence of GVHD in these individuals, and infrequent DLI need.

*Donor derived T cell recovery kinetics*

To identify the cell subset associated with lymphoid expansion and to determine the effect of donor derived T cell recovery kinetics on clinical outcomes, the *dd*CD3 counts were plotted as a function of time for the first six months after SCT for the patients for whom data was available. Utilizing the equation for the resulting curves, the derivative equation of each curve was then calculated to allow the determination of an approximate, instantaneous rate of change of *dd*CD3 at various times along the plotted curve (Figure 3B). The derivative of this curve for each patient $dx/dy$, was calculated for day 30 and 45 post-SCT in order to assess the impact of early T cell reconstitution on clinical outcomes. The $dx/dy$ values obtained in this ATG conditioned set of patients for day 30 and day 45 were highly correlated with each other (Spearman's correlation coefficient, $R^2$= 0.91). They also correlated with the logistic patterns of ALC recovery observed in each patient, suggesting that the exponential expansion observed in the ALC may be due to *dd*CD3 proliferation. Both the $dx/dy$ *(30)* and $dx/dy$ *(45)* medians are significantly different between logistic classifications (Table 3), specifically patients with pattern A logistic lymphoid recovery, had higher $dx/dy$ *(30)* compared to those demonstrating both pattern B (p-value = 0.0014) and pattern C (p-value =<0.001). For the $dx/dy$ *(45)* pattern C patients had significantly lower levels than both pattern A  (p-value 0.0063) and B patients (p value 0.0063) (using Bonferroni-adjusted significance level of 0.0167). This suggests that the exponential expansion observed in the lymphocyte recovery patterns was attributable to *dd*CD3 proliferation. And even though they are *instantaneous* measures of *dd*CD3+ cell expansion, the day 30 and 45 *dd*CD3 derivative measurements generally correlated with clinical outcomes (Table 4). This measure at day 30 is significantly different between patients experiencing GVHD (p-value = 0.0026) and those requiring DLI (p-value = 0.0476), indicating the influence of early donor T cell expansion on alloreactivity.

**Discussion.**

Patients undergoing HLA matched allogeneic SCT are at risk for competing hazards of relapse, non-relapse mortality and alloreactivity i.e. GVHD or graft rejection, potentially leading to treatment failure [32]. Beyond tumor biology, the two most critical factors influencing treatment failure are: i) regimen related toxicity of the SCT; and ii) T cell reconstitution following transplantation with its effects on alloreactivity and opportunistic infections. Once regimen related toxicity is diminished using RIC regimens, the ability to forecast, and manage the restoration of T cell immunity would be of great value in terms of optimizing clinical outcomes. In this paper we find that T cell reconstitution following SCT is a dynamic process, which can be modeled by equations commonly used to describe growth in biological systems. Further the dynamics of T cell reconstitution have a major impact on important clinical outcomes such as GVHD and engraftment as well as survival. This suggests that the likelihood of certain clinical outcomes may be computed in real time during the course of stem cell transplantation and appropriate therapeutic adjustments made dynamically, rather than at specific time points post transplant.

Systems following logistic dynamics, modeling constrained growth, depicted by the sigmoid curve, are common in biology [22, 23], describing phenomenon as diverse as population growth (in both the microbial and in the animal kingdom) and enzyme reaction kinetics with saturation. The hypothesis that they should also describe lymphocyte reconstitution following an immuno-ablative SCT is rational and supported by our data. Further support for this concept comes from the observation that as expected, clinical events such as GVHD and infections, and relapse lead to a departure from steady state behavior evident in these curves. An important consideration in reviewing the ALC data presented here is that many different immune cell subsets constitute these curves, NK cells, T cell subsets, and B cells, with all the cell subsets following different growth kinetics, so the ALC growth curves seen here are an average of all those influences. Correlation of these growth curves with donor derived T cell counts however makes it evident that this constitutes the dominant immune cell population leading to the *late exponential growth* (in the second month following SCT with this regimen) and its subsequent influence on long term clinical outcome, whether it be GVHD after withdrawal of immunosuppression or long term disease control. Once again supporting this idea is the observation of eventual relapse, and DLI need in patients who did not experience this late lymphoid/ddCD3+ cell expansion.

There are important clinical implications of this dynamical system modeling of SCT, the most important one being that it identifies subgroups of patients at greater risk for certain transplant outcomes over others (relapse or GVHD), increasing the predictability of the clinical course likely to be observed in individual SCT recipients. For instance, patients with logistic pattern A, or with a high rate of change of *dd*CD3 at day 30 in our two cohorts were more likely to develop late onset acute or chronic GVHD when immunosuppression was withdrawn. Patients with logistic pattern C were more likely to require DLI or to relapse. This behavior suggests that plotting ALC and/or *dd*CD3 in the first few weeks following SCT may allow categorization of patients into risk groups based on dynamic immune reconstitution. Future interventions may then be planned accordingly. This may include prolonged immunosuppression with tacrolimus in those with high derivative ddCD3 values, or closer monitoring and pre-emptive therapy for cancer relapse in those with lower values.

Certain dynamical systems demonstrate strong influence of early conditions on late outcomes. To understand this, consider an idealized situation where the natural killer cell and B cell effect on clinical outcome is negligible, and clinical outcomes, such as GVHD, GVM (considering standard risk malignancy in a minimal residual disease state) and engraftment are a function of T cell reconstitution. Here, the T cell population, $N_{t+1}$, at time, $t+1$ following SCT, is a function of the T cell population $N_t$, at an earlier time point, $t$. This in turn will be dependent on the T cell population at the outset of transplant $N_0$, and the growth rate $R$ governing the growth, such that

$$N_0 \xrightarrow{[R]} N_t \xrightarrow{[R]} N_{t+1} \xrightarrow{[R]} K$$

In this model, (Figure 4) the final steady state population of T cells, $K$, and its clonal repertoire will have a major influence on clinical outcomes. An approximation of this growth rate for donor derived T cells at various times following SCT is provided by the derivative we have calculated. In this model initial conditions (i.e. $N_0$ and $R$) impact the trajectory of lymphoid recovery following SCT. The higher rate of mixed T cell chimerism and lower *dd*CD3 cell count observed several months after the transplant in the ATG 7.5 cohort is consistent with this initial effect. Although the population differences in clinical outcomes are not statistically significant between the two cohorts, it is noteworthy that an ATG dose difference of 2.4 mg/kg may have an influence on immune reconstitution and subsequent clinical outcomes, reflected by absence of late acute GVHD and a greater need for DLI, as well as graft loss in three individuals conditioned with the higher ATG dose. Presumably this results from a higher ATG level at the time of cell infusion leading to lower *in vivo* $N_0$, and also potentially a lower value of $R$ in these patients. Further, the expansion of *dd*CD3+ cells would be in part dependent on antigen presentation, with T cell clones presented with antigen more likely to proliferate rapidly. Thus in patients with more robust hematopoietic recovery it is likely that the average value of $R$ would be greater as antigen presenting cell populations reconstitute faster. The correspondence of early monocyte recovery with lymphoid recovery patterns observed supports this hypothesis, in fact this may represent an example of *interacting dynamical systems*, where the $R$ value in one drives the same in the other system. Thus the $R_{antigen\ presenting\ cell}$ value may drive the $R_{effector\ T\ cell}$, by altering the cytokine concentrations in the post-transplant milieu, with ambient immunosuppression impacting both the systems.

To understand the cellular basis of the relationship between immune cell reconstitution kinetics and clinical outcomes (GVHD, relapse), it is important to recognize that the ALC curves we have obtained, as well as the ddCD3 curves represent an average measure of the proliferation of thousands of different T cell clones, when defined by T cell receptor specificity. These T cell clones are expected to follow growth curves similar to this average growth curve recorded in each individual, but with different parameters. The variation in $R$ and $N_0$ parameters will likely depend on antigen presentation, the state of immunosuppression and the inflammatory milieu at the time of transplant. When considering these parameters as variables in the logistic growth equations, the direct implication is that the clones which are poised to expand at the threshold of the exponential expansion (e.g., at the time of stopping MMF in this regimen), grow rapidly and determine whether a patient gets GVHD or not. In this scenario, if there are many proliferating alloreactive T cell clones, GVHD ensues, but if tolerance was engendered in the very first few days after transplant, GVHD will be less likely to develop. In other words, the conditions in the earliest days after transplantation determine long-term outcomes if the system' is left at steady state. Recently a similar observation has been made in murine transplant models where T cell reactivity to 3[rd] party antigens is maintained, even while host-specific alloreactivity is eliminated by administering cyclophosphamide (Cy) on day 3 and 4 post allografting [33]. The Cy works by eliminating the rapidly proliferating alloreactive T cell clones, whilst sparing the non-cycling 3[rd] party reactive clones. An important consideration in the dynamical immune system theory is that if the system is perturbed, for instance by either; i) withdrawal of stable immunosuppression ($R$ changes); or ii) the administration of DLI (new clones introduced, or $N_0$ changes); or iii) administration of vaccines with cross-reactivity to allo-antigens: GVHD may ensue. Another important property of the dynamical system described here is that, at very low growth rate populations (T cell clones in this case) decline to extinction, while at very high growth rates *chaotic* behavior with very high variability in magnitude of the populations over time may be observed. Such chaotic behavior may explain the apparent randomness in incidence of alloreactivity over time, specifically GVHD occurrence in HLA matched and mismatched SCT.  The observation of fractal organization of the T cell repertoire in normal donors and following SCT supports the notion of chaotic dynamics at work in this model [26].

The quantitative immune reconstitution model described here is well known, with the familiar sigmoid growth kinetics of lymphocytes and by extension T cell subsets and NK cells, and the salutary effect of early lymphoid recovery on clinical outcomes after SCT [15, 34]. The novelty of this work is in that it formalizes the understanding of the nature of immune recovery post transplant as a dynamical system by clearly demonstrating an exponential expansion of lymphocytes, which occurs at variable rates. In part this observation was made possible by the use of ATG as a component of this conditioning regimen, which by controlling alloreactive T cell expansion in the earliest days following SCT may allow a clearer picture of lymphoid reconstitution to emerge in some patients. Considering the large alloreactivity potential that exists between donors and recipients [19, 20], T cell-replete allografts may result in a massive proliferation of donor T cells not tolerant of recipient minor histocompatibility antigens. In these patients the oligoclonal T cell proliferation will be observed after allografting and there will be high likelihood of GVHD [35]. According to the logistic growth model, it will be the antigen exposure in the very first days following transplant, which will set the reconstituting T cell repertoire on this GVHD prone trajectory. On the other hand in grafts where this initial proliferation is controlled, either by pre-transplant ATG, or post-transplant Cytoxan or bortezomib; non-alloreactive T cell clones will dominate in the early days of transplant and will expand, out-competing alloreactive clones, and setting the system on a tolerant trajectory [26].

This dynamical system modeling differs from stochastic modeling of transplant outcomes in *individuals*, because it allows associations between immune reconstitution kinetics and clinical endpoints (either freedom from GVHD or not). In practice this would imply that immune cell subsets are measured frequently in the first few weeks after transplantation in each SCT recipient. Rate of lymphoid reconstitution at different times following SCT may then allow an estimation of the likelihood of alloreactivity developing in that specific patient, and the corresponding clinical outcome. The probability of the desired outcome, $P$ at time $t+1$, is dependent on all the *states* of the system leading up to the time $t$, preceding $t+1$; as explained above, the trajectory of the system. This way clinical outcomes may be projected based on the lymphocyte recovery kinetics observed in the first few weeks following SCT. On the other hand in a stochastic model the probability $P$ of the desired outcome at time $t+1$ is a function of the probability distribution of the health state being observed at time $t$ and is unaffected by states of the system leading up to that time [36]. Therefore an important advantage of dynamical system modeling is that by active measurement and intervention, this allows therapies to be modified on an ongoing basis to achieve desired clinical outcomes. On the other hand, conventional stochastic modeling allows selection of the optimal therapeutic program, but once the treatment is initiated, can only predict the outcomes observed based on prior probability distributions, allowing only empiric therapeutic modifications. Thus dynamical system modeling of immune reconstitution would allow true personalization of patient management. Eventually integrating this with tumor growth kinetics should allow further improvements in clinical management of patients.

In conclusion, we demonstrate that immune reconstitution following SCT may be modeled mathematically using equations that are commonly used to describe population growth. Modeling immune reconstitution following other conditioning and GVHD prophylaxis regimens will allow determination of the degree to which such measurement may be generalizable and as such allow real time adjustments to be made in post-transplant immunosuppression resulting in optimal immunosuppression. Importantly, dynamical system modeling of transplant makes it possible to limit the effect of randomness in predicting clinical outcomes such as GVHD and engraftment and further refine trial design by accounting for immune reconstitution kinetics.

**Acknowledgements:** The authors gratefully acknowledge Sanofi-Aventis, the manufacturers of Thymoglobulin for their support of this clinical trial. The authors also gratefully acknowledge Ms. Carol Cole, Laura Couch, Judith Davis, Angela Buskey, Dana Broadway and Kathryn Candler for their excellent help in executing this protocol and the care of the patients enrolled on this study. We also acknowledge the Virginia Commonwealth University BMT program's nurse practitioners and nurses for their contributions towards the completion of this study. Dr. Neale was supported by Commonwealth Health Research Board grant #236-11-13.

**Table 1**. Patient characteristics.

|  | ATG 5.1 | ATG 7.5 |
|---|---|---|
| N | 19 | 22 |
| Male | 12 | 14 |
| Median age (range) | 57 (44-69) | 57 (40-68) |
| Donor |  |  |
| MRD | 9 | 10 |
| URD | 10 | 12 |
| HLA Mismatch | 2 | 2 |
| Stem cell source |  |  |
| Bone Marrow | 2 | 2 |
| Peripheral Blood | 17 | 20 |
| Diagnosis and prior therapy |  |  |
| MM | 5 | 4 |
| NHL | 7 | 8 |
| AML | 1 | 3 |
| MDS | 0 | 2 |
| CLL | 4 | 3 |
| PLL | 2 | 2 |
| Median # prior regimens | 4 (2-10) | 4 (1-15) |
| Prior Auto SCT | 6 | 8 |
| Median cell dose Infused |  |  |
| CD3+ | 2.3 (0.1-5.7) | 2.9 (0.2-11.3) |
| CD34+ | 5.8 (1.7-10.4) | 5.1 (1.6-7.5) |

**Table 2:** Median percentage recipient chimerism (top) and *dd*CD3+ cell counts (bottom) by ATG dosing cohort and time post SCT in months.

|  | **1** | **2** | **3** | **6** | **9** | **12** |
|---|---|---|---|---|---|---|
| **ATG 7.5** | 6.7 | 8.8 | 18.5 | 8.7 | 0.0 | 1.4 |
| **ATG 5.1** | 0.0 | 1.0 | 1.5 | 0.0 | 0.0 | 0.0 |
| *p-value on difference* | *0.20* | *0.16* | *0.024\** | *0.056* | *0.052* | *0.077* |
| **ATG 7.5** | 93.3 | 105.8 | 92.1 | 225.7 | 352.0 | 493.0 |
| **ATG 5.1** | 214.1 | 576.6 | 405.0 | 408.0 | 853.7 | 680.0 |
| *p-value on differences* | *0.068* | *0.073* | *0.031\** | *0.63* | *0.071* | *0.18* |

**Table 3:** Correlation of rate of change in *dd*CD3+ cell counts at days 30 and 45 with ALC recovery logistic patterns (Medians and IQR)

| Logistic pattern | *dx/dy (30)* **dd**CD3 | | | *dx/dy (45)* **dd**CD3 | | |
|---|---|---|---|---|---|---|
|  | N | Median | IQR | N | Median | IQR |
| **A** | 14 | 15.5 | 12.2, 25.2 | 14 | 10.8 | 1.4, 17.3 |
| **B** | 14 | 2.5 | 0.4, 3.9 | 14 | 1.5 | 0.6, 26 |
| **C** | 5 | 0.2 | -0.1, 0.5 | 5 | 0.1 | -0.1, 0.5 |
|  | *p-value < 0.0001* | | | *p-value = 0.0015* | | |

**Table 4:** Correlation of day 30 and day 45 *dd*CD3 derivative values and clinical outcome.

| *dx/dy(30)* | | | | | | | |
|---|---|---|---|---|---|---|---|
| | **Yes** | | | **No** | | | |
| | N | Median | IQR | N | Median | IQR | *p-value* |
| **GVHD** | 16 | *14.5* | 4.4, 16.5 | 17 | *0.9* | 0.1, 5.5 | ***0.0026*** |
| **Relapse** | 14 | *2.3* | -0.1, 10.9 | 19 | *4.7* | 2.1, 16.0 | *0.0869* |
| **Survival** | 22 | *4.5* | 0.9, 15.6 | 11 | *3.7* | 0.3, 15.5 | *0.8635* |
| **DLI Use** | 9 | *0.9* | 0.1, 5.2 | 24 | *7.1* | 1.4, 16.5 | ***0.0476*** |
| *dx/dy(45)* | | | | | | | |
| **GVHD** | 16 | *3.4* | 1.2, 13.5 | 17 | *0.7* | 0.3, 2.3 | *0.0562* |
| **Relapse** | 14 | *1.6* | 0.4, 6.9 | 19 | *1.8* | 0.9, 10.9 | *0.4888* |
| **Survival** | 22 | *1.6* | 0.8, 6.5 | 11 | *1.9* | 0.1, 15.6 | *0.9848* |
| **DLI Use** | 9 | *0.5* | 0.3, 2.0 | 24 | *2.1* | 0.9, 13.2 | *0.1149* |

**Figure 1. Clinical outcomes in the ATG 5.1 and ATG 7.5 cohorts. (A)** Kaplan-Meier Survival curves. **(B)** Cumulative incidence curves for relapse (competing risk of TRM). **(C)** Kaplan-Meier Event-Free Survival (relapse and fatality). (D) Cumulative incidence curves for chronic GVHD, observations censored at DLI (competing risk of fatality without GVHD). (E) Cumulative incidence curve for DLI (competing risk of fatality)

**A.**

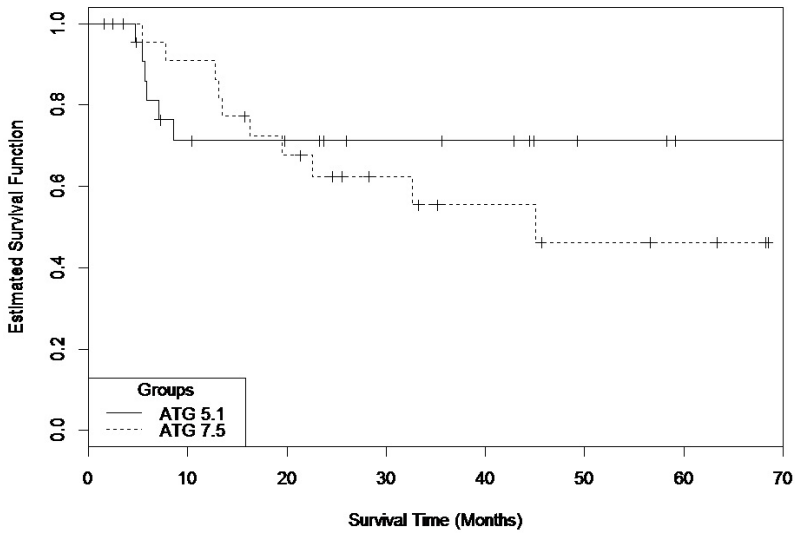

**B.**

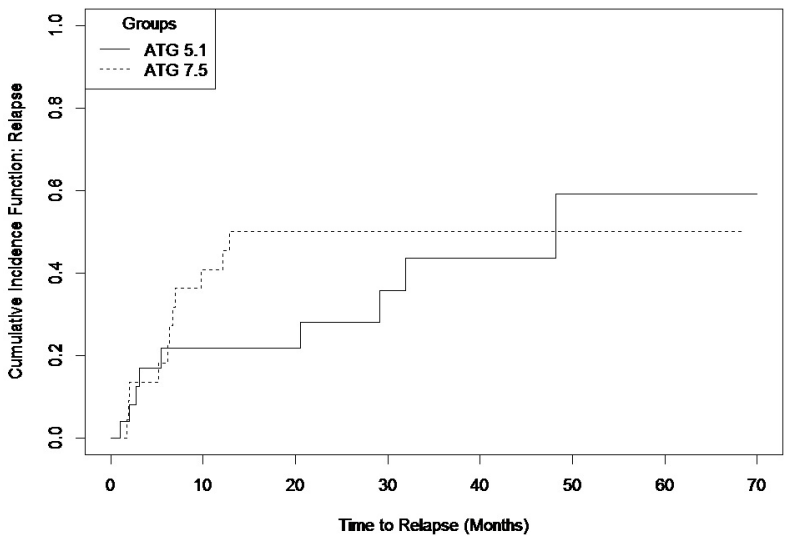

**C.**

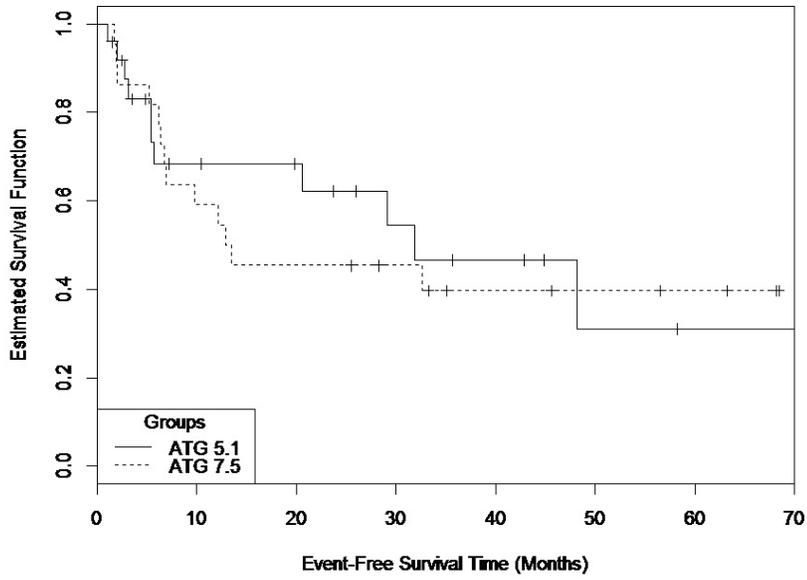

**D.**

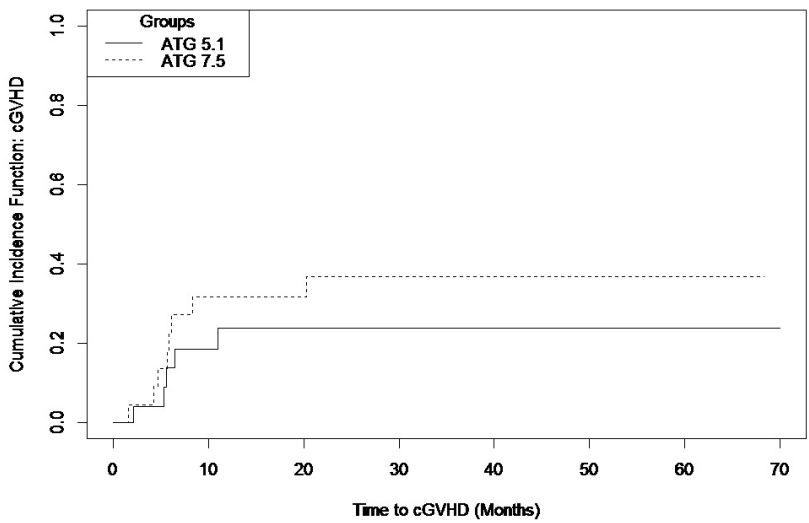

**E.**

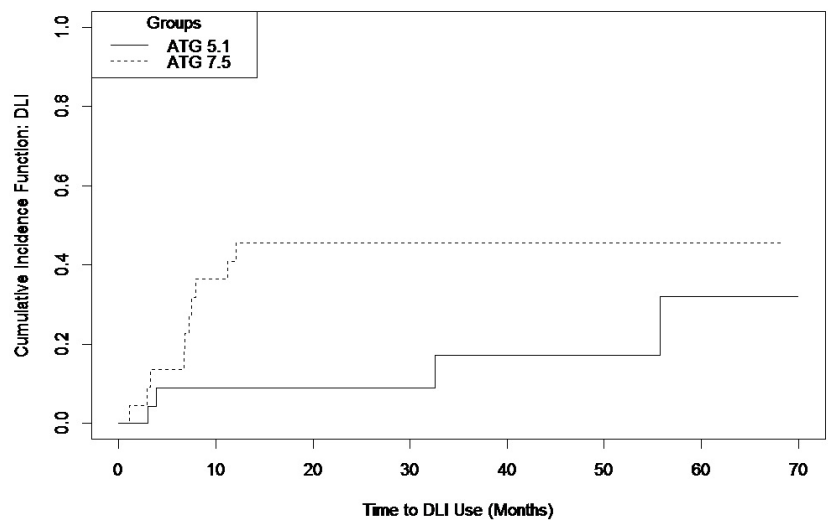

**Figure 2. Absolute lymphocyte count (μL$^{-1}$) recovery following ATG+TBI and SCT**. **(A)** Two periods of growth are seen, first from day 0-30 when patients are on MMF (blue outline), and the second, from day 20 to approximately day 120 (orange outline). Each of these periods demonstrates an initial phase of slow growth, followed by exponential expansion and finally relatively stable counts, until tacrolimus is discontinued, when greater variability is observed. **(B)** Logistic dynamics of lymphoid recovery following SCT. Three patterns are observed corresponding to high, pattern A, intermediate, pattern B and slow growth rate, pattern C.

**A.**

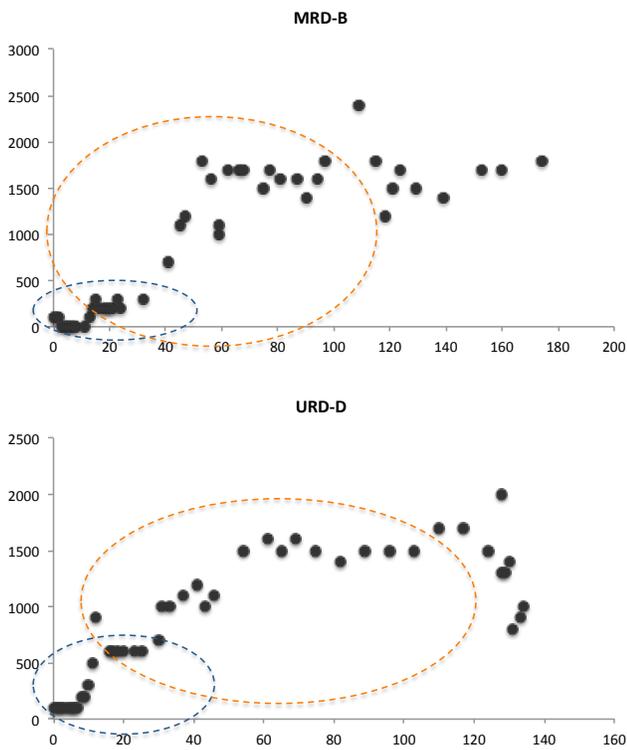

**B.**

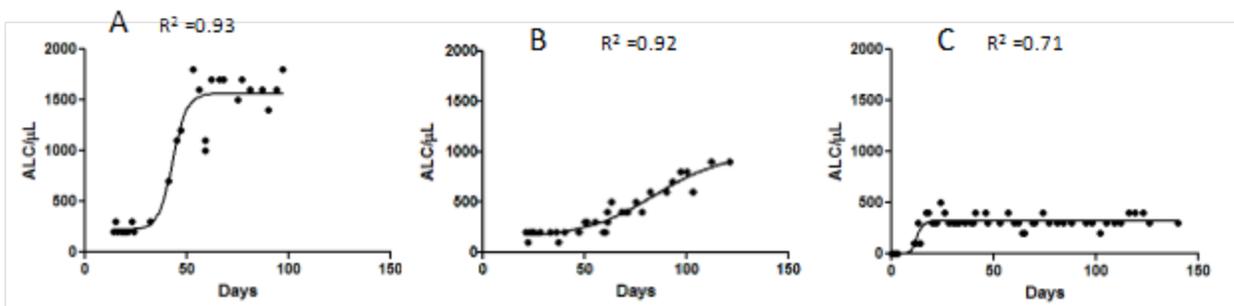

**Figure 3. Determinants of lymphoid recovery.** (A) Correspondence of early monocyte recovery with lymphocyte reconstitution, association of high early monocyte recovery with late lymphoid expansion (orange circles- monocytes; black diamonds lymphocytes, both µL$^{-1}$) plotted against days post SCT. (**B**) Donor derived CD3+ cell count plotted over time, and derivative for the polynomial curve determined at day 30 and day 45. Derivative is the slope of the tangent to the polynomial curve (*dy/dx*), and gives the approximate rate of change of *dd*CD3 at that point in time. Accuracy of the derivative will depend on the goodness of fit of the polynomial curve.

**A.**

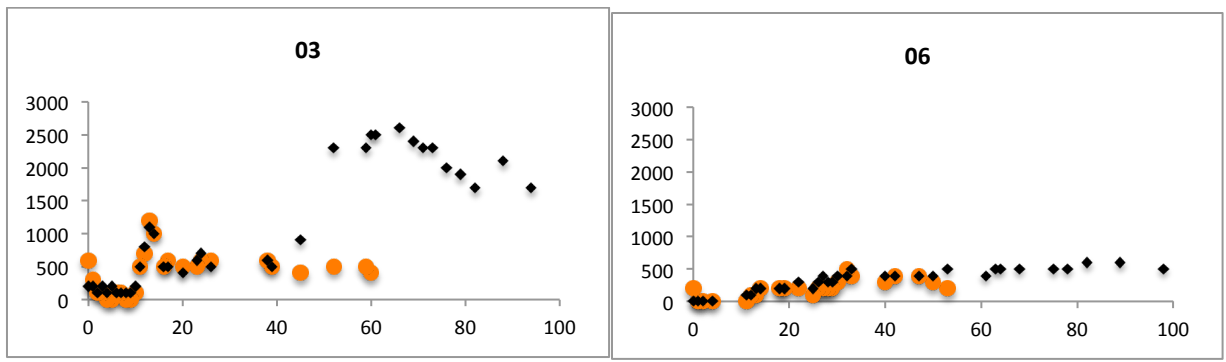

**B.**

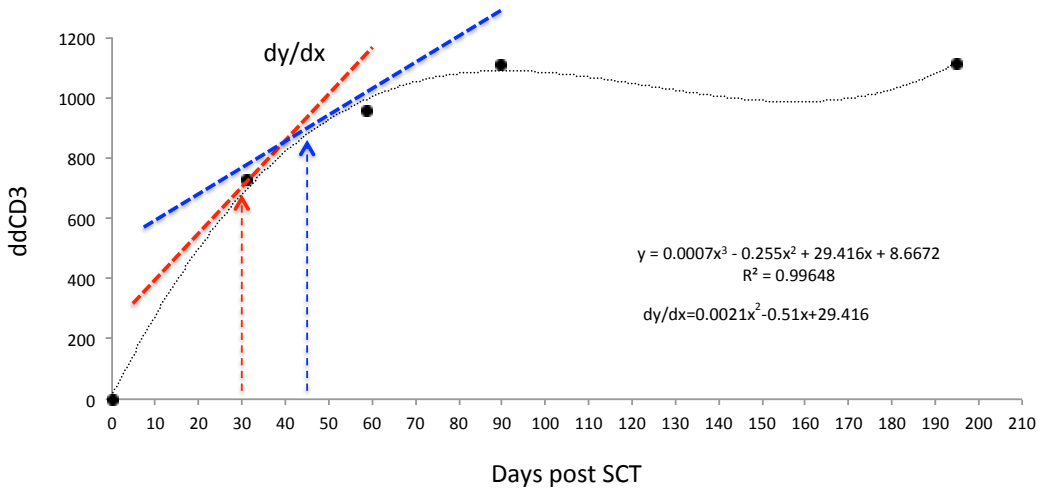

**Figure 4.** Logistic model of lymphocyte proliferation post transplant. Cell dose infused at the time of SCT $N_0$, will start proliferating under the influence of the driving parameter $R$ and expand over time $t$, taking on values of $N_t$, $N_{t+1}$...., $N_{t+n}$. This will include a period of exponential expansion as the lymphopenic state following SCT is corrected and eventually a stable lymphocyte population $K$ is established. Changes in antigen presentation (due to infection, tissue injury) and changes in immunosuppression levels and the overall cytokine milieu may perturb the steady state or growing populations by altering $R$ and result in greater variability seen at specific time points. Reconstitution of the lymphocyte subset populations by differentiation of hematopoietic cells also changes $N$. The curves represents an average of the lymphoid cell populations (and clones) seen in circulation, with each population repopulating with similar dynamics, but at different rates. Three other hypothetical curves along the continuum of logistic expansion are also shown with different $N_0$ (not evident due to scale) and $R$ values ($R>R1>R2>R3$). Blue circles and top curve depict data from an actual patient.

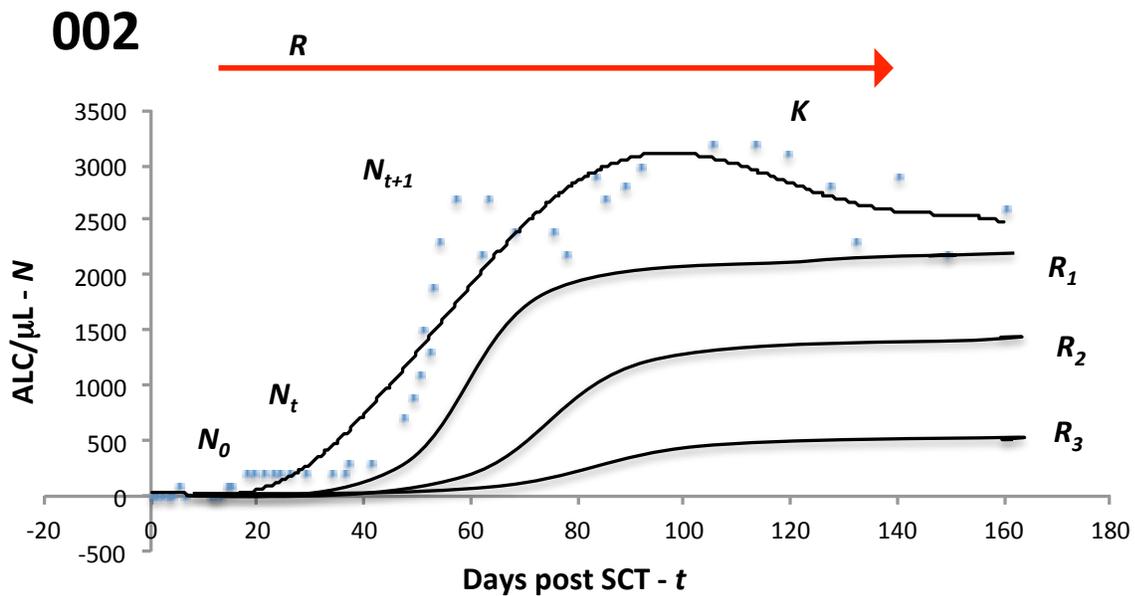

**Supplementary Figure 1**. Schema of the transplant protocol, outlining the general immunosuppression withdrawal scheme.

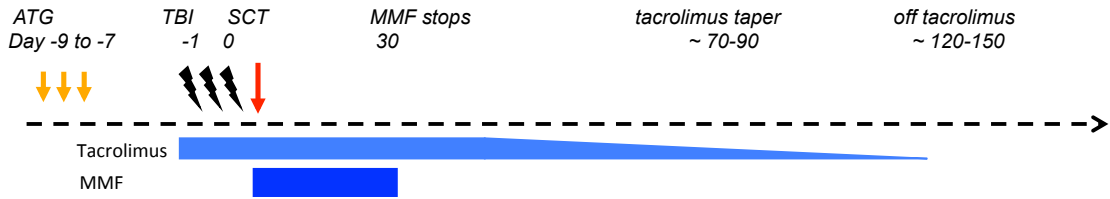

**Supplementary Figure 2**. Mean T cell chimerism and *dd*CD3 plotted over time since transplant in the ATG 5.1 and ATG 7.5 cohort.

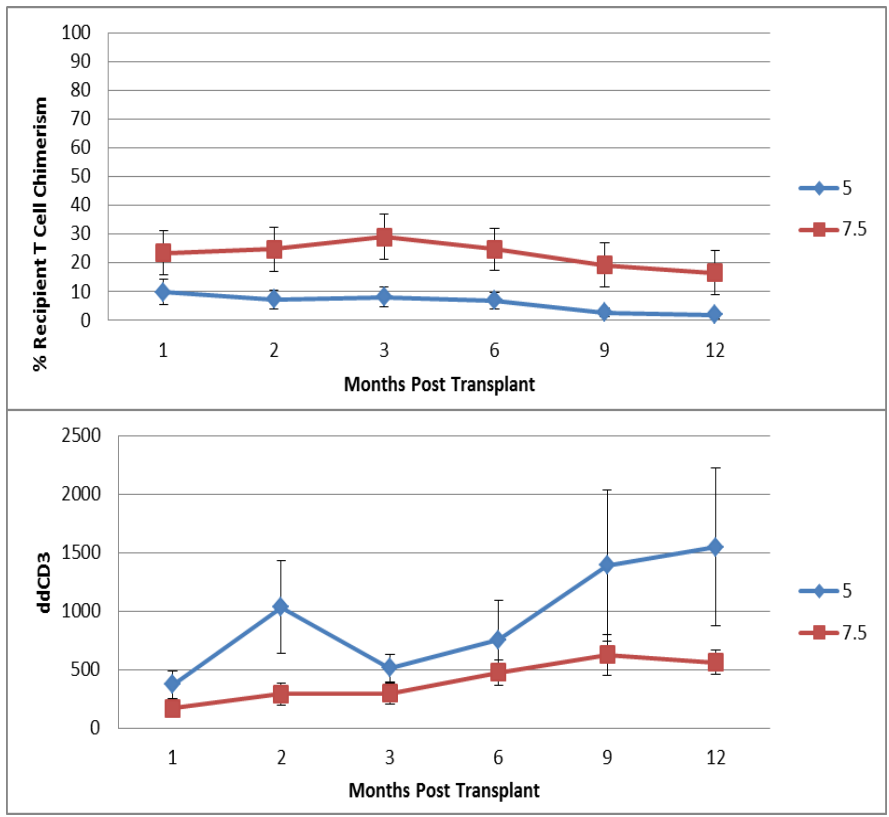

*Mean, SE

**Supplementary Figure 3**. Three logistic patterns evident in patients conditioned with ATG + 4.5 Gy TBI, based on peak ALC and/or time to peak (cross-hairs).  Pattern A, high, rapid, ALC recovery; Pattern B, intermediate, slower recovery; Pattern C, poor ALC recovery. X- axis (days post SCT) and Y-axis (ALC, µl$^{-1}$) are all scaled to the approximately same value except for patient #035 in pattern A.  ALC recovery graph for 3 patients in the ATG 7.5 group not shown (predominantly recipient T cell chimeric, before day 100).

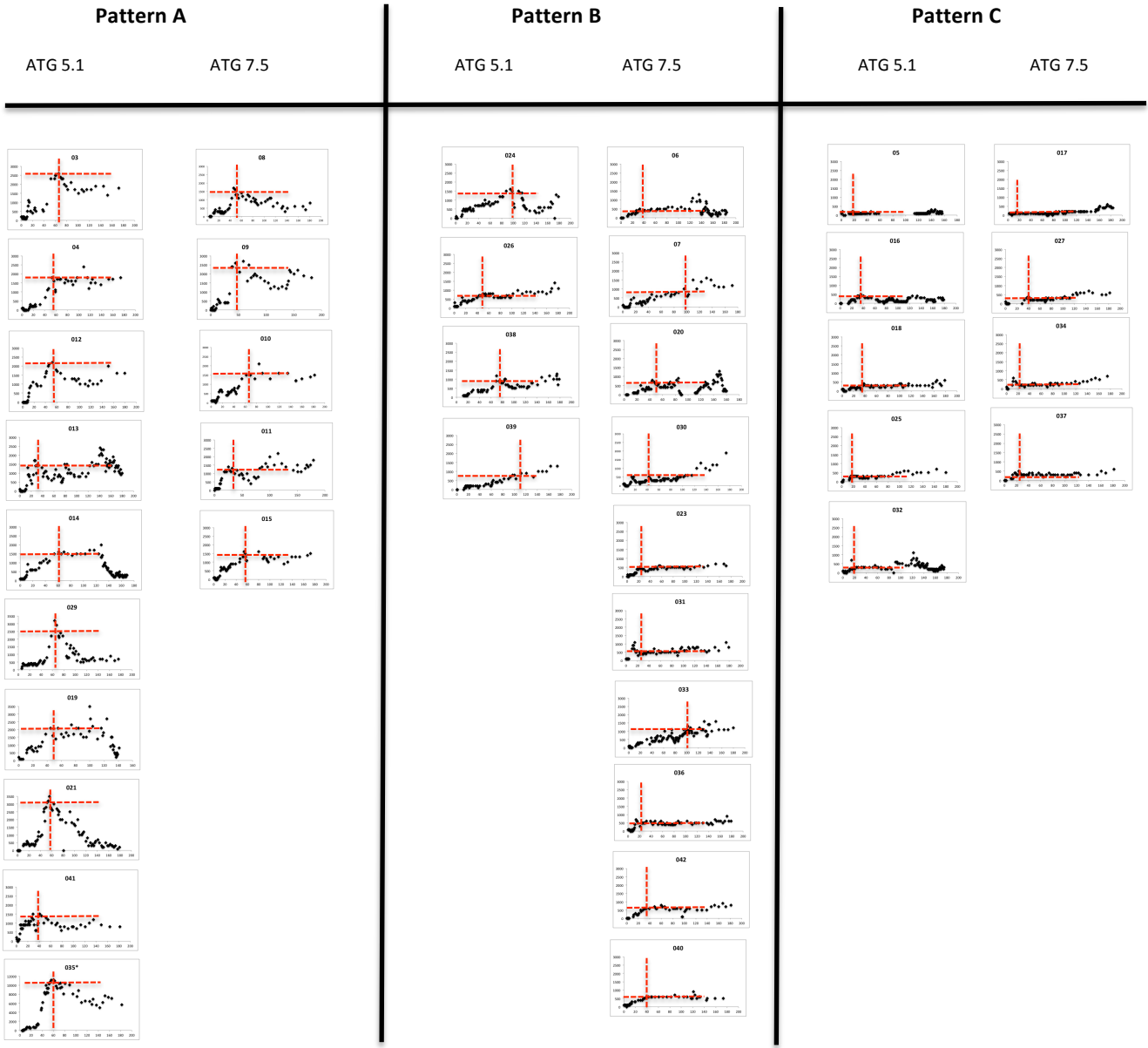

**Supplementary Figure 4**. Absolute lymphocyte counts (ALC) and absolute monocyte counts (AMC) for individuals over time following SCT. Correspondence of post engraftment AMC with ALC expansion observed later in the course in patients demonstrating different logistic lymphocyte growth patterns (A, B and C).

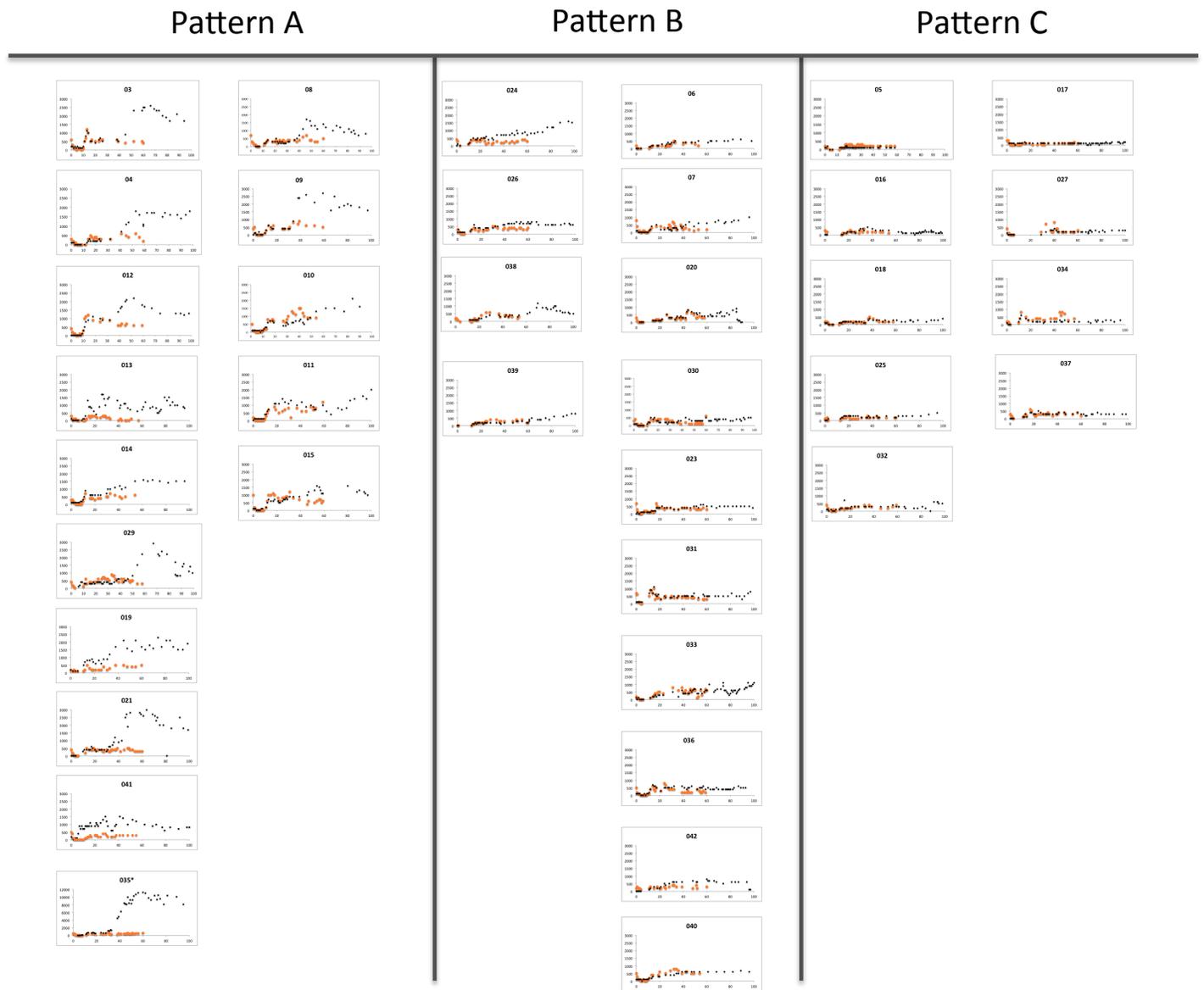

**Supplementary Methods.**

*Supportive care and GVHD prophylaxis*

All patients received oral levofloxacin 500 mg daily from day -9 to neutrophil engraftment, valacyclovir 1000 mg three times daily, voriconazole 200 twice daily and sulfamethoxazole-trimethoprm DS twice daily on two days each week from day -9 till taken off tacrolimus following which voriconazole was discontinued and valacyclovir dose was reduced to twice daily. Methylprednisolone 1 mg/kg twice daily was administered on the days of ATG infusion, along with diphenhydramine and acetaminophen.

Tacrolimus levels were maintained at 10-14 ng/mL in the first month following transplant, following which the levels were generally maintained in the 8-12 ng/mL range until day 90 to 120, following which dose was tapered. The area under the plasma concentration time curve (AUC) for tacrolimus was calculated using the linear trapezoidal method for approximating numerical integrals. All documented serum tacrolimus trough levels (i.e. collected within 1 hour of the next scheduled dose) from day 0 through day +90 were collected and included in the estimation of the AUC. The trapezoidal AUC was calculated using the equation

$$\Delta \mathrm{AUC}_{1\text{-}2} = \frac{(\mathrm{Cp}_1 + \mathrm{Cp}_2)}{2} \times (t_2 - t_1)$$

where $Cp_1$ and $Cp_2$ represent tacrolimus concentrations at times $t_1$ and $t_2$, respectively. The sum total AUC was then calculated for each patient at the following time points: day +15, +30, and +60.